\input harvmac
\input amssym.def
\input amssym.tex

\def\tablerule{\noalign{\hrule}}

\parskip=4pt \baselineskip=12pt
\hfuzz=20pt
\parindent 10pt

\def\tcc{{\tilde{\cal C}}}
\def\hh{{\tilde h}}
\def\mod{{\rm mod}}
\def\n{{\nu}}
\def\fha{{\textstyle{5\over2}}}
\def\han{{\textstyle{n\over2}}}

\def\newsubsec#1{\global\advance\subsecno by1\message{(\secsym\the\subsecno.
#1)} \ifnum\lastpenalty>9000\else\bigbreak\fi
\noindent{\bf\secsym\the\subsecno. #1}\writetoca{\string\quad
{\secsym\the\subsecno.} {#1}}}

\global\newcount\subsubsecno \global\subsubsecno=0
\def\subsubsec#1{\global\advance\subsubsecno
by1\message{(\secsym\the\subsecno.\the\subsubsecno. #1)}
\ifnum\lastpenalty>9000\else\bigbreak\fi
\noindent{\bf\secsym\the\subsecno.\the\subsubsecno.
#1}\writetoca{\string\quad
{\secsym\the\subsecno.\the\subsubsecno.}
{#1}}\par\nobreak\medskip\nobreak}

\input epsf.tex
\newcount\figno
\figno=0
\def\fig#1#2#3{
\par\begingroup\parindent=0pt\leftskip=1cm\rightskip=1cm\parindent=0pt
\baselineskip=11pt \global\advance\figno by 1 \midinsert
\epsfxsize=#3 \centerline{\epsfbox{#2}} \vskip 12pt
#1\par
\endinsert\endgroup\par}
\def\figlabel#1{\xdef#1{\the\figno}}
\def\encadremath#1{\vbox{\hrule\hbox{\vrule\kern8pt\vbox{\kern8pt
\hbox{$\displaystyle #1$}\kern8pt} \kern8pt\vrule}\hrule}}

  \def\tV{{\tilde V}}
\def\tcn{{\tilde{\cal N}}}
\def\newsubsubsec#1{\global\advance\subsubsecno
by1\message{(\secsym\the\subsecno.\the\subsubsecno. #1)}
\ifnum\lastpenalty>9000\else\bigbreak\fi
\noindent{\bf\secsym\the\subsecno.\the\subsubsecno.
#1}\writetoca{\string\quad {\secsym\the\subsecno.\the\subsubsecno.}
{#1}}}

\def\nt{\noindent}
\def\nl{\hfill\break}

\def\np{\vfill\eject}

\def\rank{{\rm rank}}

\def\riga{-\kern-4pt - \kern-4pt -}
\font\fat=cmsy10 scaled\magstep5

\def\Bbullet{\raise-3pt\hbox{\fat\char"0F}}

\font\tfont=cmbx12 scaled\magstep1 
\font\male=cmr9

\def\Box{
\vbox{ \halign to5pt{\strut##& \hfil ## \hfil \cr &$\kern -0.5pt
\sqcap$ \cr \noalign{\kern -5pt \hrule} }}~}

\def\down{\raise1.5pt\hbox{$\phantom{a}_2$}\downarrow}

\def\downa{\raise1.5pt\hbox{$\phantom{a}_{2\atop m_2}$}\downarrow}

\def\({\left(}
\def\){\right)}
\def\eps{\epsilon}
 
\def\lra{\longrightarrow}

\def\lg{\langle} \def\rg{\rangle} 

\def\ha{{\textstyle{1\over2}}}

\def\trha{{\textstyle{3\over2}}}

\def\bbc{{C\kern-6.5pt I}}
\def\bac{{C\kern-5.5pt I}}
\def\bab{{C\kern-4.5pt I}}
\def\bbz{Z\!\!\!Z}

\def\bbr{{I\!\!R}}
\def\bbn{I\!\!N}
\def\a{\alpha}

\def\d{\delta}

\def\vr{\vert}

\def\l{\lambda}

\def\ca{{\cal A}}  \def\cc{{\cal C}}
\def\cd{{\cal D}} \def\ce{{\cal E}} \def\cf{{\cal F}}
\def\cg{{\cal G}} \def\ch{{\cal H}} 
 \def\ck{{\cal K}} 
\def\cm{{\cal M}} \def\cn{{\cal N}} 
\def\cp{{\cal P}}

\def\th{\theta} 

\def\idos{intertwining differential operators}

\def\L{\Lambda}
\def\r{\rho}



\lref\GeV{I.M. Gelfand, M.I. Graev and N.Y. Vilenkin, {\it
Generalised Functions}, vol. 5 (Academic Press, New York, 1966).}

\lref\GeNa{I.M. Gelfand and M.A. Naimark, Acad. Sci. USSR. J. Phys.
{\bf 10} (1946) 93-94.}

\lref\Barg{V. Bargmann, Annals Math. {\bf 48}, (1947) 568-640.}

\lref\Dobmul{V.K. Dobrev, Lett. Math. Phys. {\bf 9} (1985) 205-211.}

\lref\EHW{T. Enright, R. Howe and W. Wallach, "A classification of
unitary highest weight modules", in: {\it Representations of
Reductive Groups}, ed. P. Trombi (Birkh\"auser, Boston, 1983) pp.
97-143.}

\lref\Har{Harish-Chandra, "Discrete series for semisimple Lie
groups: II", Ann. Math. {\bf 116} (1966) 1-111.}

\lref\HC{Harish-Chandra, "Representations of semisimple Lie groups:
IV,V", Am. J. Math. {\bf 77} (1955) 743-777, {\bf 78} (1956) 1-41.}

\lref\KnSt{A.W. Knapp and E.M. Stein,
Ann. Math. {\bf 93} (1971) 489-578; II : Inv. Math. {\bf 60} (1980)
9-84.}

\lref\War{G. Warner, {\it Harmonic Analysis on Semi-Simple Lie
Groups I}, (Springer, Berlin, 1972).}

\lref\Lan{R.P. Langlands, {\it On the classification of irreducible
representations of real algebraic groups}, Math. Surveys and
Monographs, Vol.  31 (AMS, 1988), first as IAS Princeton preprint
(1973).}

\lref\Zhea{D.P. Zhelobenko, {\it Harmonic Analysis on Semisimple
Complex Lie Groups}, (Moscow, Nauka, 1974, in Russian).}

\lref\Helg{S. Helgason, {\it Differential Geometry and Symmetric
Spaces} (Academic Press, New York, 1962).}

\lref\KaVe{M. Kashiwara and M. Vergne, Invent. Math. {\bf 44} (1978)
1-47.}

\lref\Wal{N.R. Wallach,  Trans. AMS {\bf 251} (1979) 1-7 \& 19-37.}

\lref\KnWa{A.W. Knapp and N.R. Wallach, "Szeg\"o kernels associated with discrete series", Invent.
Math. {\bf 34} (1976) 163-200; ~"Correction and addition to Szeg\"o kernels
associated with discrete series", Invent. Math. {\bf 62} (1980) 341-346.}

\lref\Par{R. Parthasarathy, Proc. Indian Acad. Sci. Sect. A Math.
Sci. {\bf 89} (1980) 1–24.}

\lref\Jak{H.P. Jakobsen, Invent. Math. {\bf 62} (1980) 67-78; Math.
Ann. {\bf 256} (1981) 439-447; J. Funct. An. {\bf 52} (1983)
385-412.}

\lref\RSW{J. Rawnsley, W. Schmid and J.A. Wolf, J. Funct. Anal. {\bf
51} (1983)  1–114.}

\lref\Blank{B.E. Blank,  "Knapp-Wallach Szeg\"o integrals and
generalized principal series representations: the parabolic rank one
case", J. Funct. Anal. {\bf 60} (1985) 127-145; ~"Knapp-Wallach
Szeg\"o integrals. II. The higher parabolic rank case", Trans. Amer.
Math. Soc. {\bf 300} (1987) 49-59.}

 \lref\Mack{G. Mack, Commun.
Math. Phys. {\bf 55} (1977) 1.}

\lref\PeSo{V.B. Petkova and G.M. Sotkov, Lett. Math. Phys.
{\bf 8} (1984) 217-226; Erratum-ibid. {\bf 9} (1985) 83.}

\lref\KnZu{A.W. Knapp and G.J. Zuckerman, in: Lecture Notes in
Math., Vol. 587 (Springer, Berlin, 1977) pp. 138-159; ~Ann. Math.
{\bf 116} (1982) 389-501.}

\lref\DMPPT{V.K. Dobrev, G. Mack, V.B. Petkova, S.G. Petrova and
I.T. Todorov, {\it Harmonic Analysis on the  $n$-Dimensional Lorentz
Group and Its Applications to Conformal Quantum Field Theory},
Lecture Notes in Physics, Vol. 63  (Springer-Verlag,
 Berlin-Heidelberg-New York, 1977).}

\lref\DoPea{V.K. Dobrev and V.B. Petkova, Reports Math. Phys. {\bf
13} (1978) 233-277.}

\lref\Dirac{P.A.M. Dirac, J. Math. Phys. {\bf 4}, 901 (1963).}

\lref\Fro{C. Fronsdal, Rev. Mod. Phys. {\bf 37}, 221 (1965); Phys.
Rev. {\bf D10}, 589 (1974);   Phys. Rev. {\bf D12}, 3819 (1975).}

\lref\Evans{N.T. Evans, J. Math. Phys. {\bf 8}, 170 (1967).}

\lref\FF{M. Flato and C. Fronsdal, Lett. Math. Phys. {\bf 2},
421-426 (1978); Phys. Lett. {\bf B97}, 236 (1980); J. Math. Phys.
{\bf 22} , 1100 (1981); Phys. Rev. {\bf D23}, 1278 (1981); Phys.
Scripta {\bf 24}, 895 (1981).}

\lref\Dobc{V.K. Dobrev,
J. Math. Phys. {\bf 26} (1985) 235-251.}

\lref\Knapp{A.W. Knapp, {\it Representation Theory of Semisimple
Groups (An Overview Based on Examples)}, (Princeton Univ. Press,
1986).}

\lref\Dob{V.K. Dobrev,
Rep. Math. Phys. {\bf 25} (1988) 159-181; first as ICTP Trieste
preprint IC/86/393 (1986).}

\lref\Dobsrni{V.K. Dobrev,   Suppl. Rendiconti Circolo Matematici di
Palermo, Serie II, Numero 43 (1996) 15-56.}

 \lref\DoMo{V.K. Dobrev  and P. Moylan,
Fort. d. Phys. {\bf 42} (1994) 339-392.}

\lref\DNW{L. Dolan, C.R. Nappi and E. Witten,
JHEP 0110 (2001) 016, hep-th/0109096.}

\lref\Witten{E. Witten, "Conformal Field Theory in Four and Six
Dimensions", arXiv:0712.0157.}

\lref\Dobso{V.K. Dobrev,
J. Phys. {\bf A39} (2006) 5995-6020; hep-th/0512354.}

\lref\Dobqg{V.K. Dobrev,   Lett. Math. Phys. {\bf 22} (1991)
251-266;   ~V.K. Dobrev and P.J. Moylan, Phys. Lett. {\bf 315B}
(1993) 292-298; ~V.K. Dobrev and R. Floreanini,   J. Phys. A: Math.
Gen. {\bf 27} (1994) 4831-4840; ~V.K. Dobrev,   J. Phys. A: Math.
Gen. {\bf 27} (1994) 4841-4857 \& 6633-6634, hep-th/9405150.}

\lref\Bru{F. Bruhat, Bull. Soc. Math. France, {\bf 84} (1956) 97-205.}

\lref\Dobcond{V.K. Dobrev, J. Phys. A: Math. Gen. {\bf 28} (1995)
7135 - 7155.}

\lref\Dobp{V.K. Dobrev, in preparation.}

\lref\DoPe{V.K. Dobrev and V.B. Petkova,
Lett. Math. Phys. {\bf 9} (1985) 287-298;~
Fortschr. d. Phys. {\bf 35} (1987) 537-572; ~
Phys. Lett. {\bf 162B} (1985) 127-132.}

\lref\Dobsu{V.K. Dobrev,
J. Phys. {\bf A35} (2002) 7079-7100, hep-th/0201076; ~V.K. Dobrev and R.B. Zhang,
Phys. Atom. Nuclei, {\bf 68} (2005) 1660-1669,
hep-th/0402039; ~V.K. Dobrev, A.M. Miteva, R.B. Zhang and B.S. Zlatev,
Czech. J. Phys. {\bf 54} (2004) 1249-1256; hep-th/0402056.}

\lref\Min{S. Minwalla,
Adv. Theor. Math. Phys. {\bf 2}  (1998) 781-846, hep-th/9712074.}

\lref\CCTV{C. Carmeli, G. Cassinelli, A. Toigo and V.S. Varadarajan,
Comm. Math. Phys. {\bf 263}  (2006) 217-258, hep-th/0501061.}


\rightline{SISSA 77/2007/EP}
\rightline{INRNE-TH-07-12}

\vskip 1.5cm

\centerline{{\tfont Positive Energy Representations,}}\vskip 2truemm
\centerline{{\tfont Holomorphic Discrete Series}}\vskip 2truemm
\centerline{{\tfont and Finite-Dimensional Irreps}}

\vskip 1.5cm

\centerline{{\bf V.K. Dobrev}}
\vskip 0.5cm

\centerline{Scuola Internazionale Superiore di Studi Avanzati}
\centerline{via Beirut 2-4} \centerline{34014 Trieste, Italy}

\vskip 0.5cm \centerline{and} \vskip 0.5cm

 \centerline{Institute for Nuclear
Research and Nuclear Energy\foot{Permanent address.}}
\centerline{Bulgarian Academy of Sciences} \centerline{72
Tsarigradsko Chaussee, 1784 Sofia, Bulgaria}

\vskip 1.5cm

 \centerline{{\bf Abstract}}
\midinsert\narrower{\male Let ~$G$~ be a semi-simple non-compact Lie
group with unitary lowest/highest weight representations. We
consider explicitly the relation between three types of
representations of $G$: positive energy (unitary lowest weight)
representations, (holomorphic) discrete series representations and
non-unitary finite-dimensional irreps. We consider mainly the
conformal groups $SO_o(n,2)$ treating in full detail the cases
$n=1,3,4$.}\endinsert

\vskip 1.5cm
\np

\newsec{Introduction}

\nt Let ~$G$~ be a semi-simple non-compact Lie group with unitary
lowest/highest weight representations, i.e.,   ~$(G,K)$~ is a
Hermitian symmetric pair, where $K$ is a maximal compact subgroup of
$G$ \HC. Let $\cg$ be the Lie algebra of $G$. Then, $\cg$ is one of
the following Lie algebras: $su(m,n)$, $so(n,2)$, $sp(2n,R)$,
$so^*(2n)$, $E_{6(-14)}$, $E_{7(-25)}$ \Helg. These groups/algebras
have also discrete series representations, 
since ~$\rank\, G = \rank\, K$, and ~$G \supset K\supset H$, where
~$H$~ is a Cartan subalgebra of ~$G$ \Har.

In this  paper we start a discussion on the relation between three
types of representations of $G$: positive energy (i.e., unitary
lowest weight) representations, (holomorphic) discrete series
representations and finite-dimensional representations (the latter
are not unitary).

There are  some general facts that are known about these
relationships. For example every discrete series representation has
the same infinitesimal character (Casimirs) as some
finite-dimensional representation \Har. According to the results of
\HC,\Har,\EHW{} a holomorphic/antiholomorphic discrete series
representation is (infinitesimally) equivalent to a unitary
lowest/highest weight representation.
The   submerging of the set of discrete points enumerating the
holomorphic discrete series in the semi-infinite interval
parametrizing the unitary lowest weight representations is given in
\EHW.\foot{Note that EHW \EHW{} work in the conjugate picture with
highest weight modules.}  The embedding of discrete series
representations into elementary representations\foot{The precise
definition with relevant references is given below.} (called also
generalized principal series representations) is given in
\KnWa,\Knapp,\Blank. (Other pertinent references are
\KaVe,\Wal,\Par,\Jak,\RSW,\DNW.) However, the latter two very
important connections are not used simultaneously in the
mathematical literature.

Our input in  the discussion of these relationships may be
summarized as follows. First of all, we use all relationships
between the three types of representations. In particular, we use
essentially the fact that discrete series representations and
finite-dimensional representations occur as subrepresentations of
elementary representations. The elementary representations in
question are topologically reducible (and not unitary). We group the
(reducible) ERs with the same Casimirs  in sets called ~{\it
multiplets} \Dobmul,\Dob{}. The multiplet corresponding to fixed
values of the Casimirs may be depicted as a connected graph, the
vertices of which correspond to the reducible ERs and the lines
between the vertices correspond to intertwining operators.\foot{For
simplicity only the operators which are not compositions of other
operators are depicted.  The ERs which are related by non-trivial
intertwining operators are said to be partially equivalent.} The
explicit parametrization of the multiplets and of their ERs is
important for understanding of the situation. Especially important
are the multiplets containing (as subrepresentation of some
reducible ER of the multiplet) a finite-dimensional representation.
Each such multiplet contains some discrete series representation(s)
(as subrepresentation(s) of other reducible ER(s) of the multiplet),
and all discrete series representations are contained in some such
multiplet. Furthermore, from these multiplets - by certain limiting
procedure - one can obtain all multiplets containing limits of
discrete series. (The latter resulting multiplets   do not contain
finite-dimensional representations.)  Finally, using the multiplets,
we can identify the intertwining operators relevant for discrete
series representations and the finite-dimensional representation.\nl
We should also mention that in distinction to mathematicians, in our
considerations we are using induction also from maximal parabolics
that are not   cuspidal (e.g., on the example of $\cg = so(n,2)$ in
this paper). 

In the present paper we start such a description by discussing in
some detail the conformal case when ~$\cg ~=~ so(n,2)$. There are
two typical cases when ~$n>2$~: ~$n$~  odd and ~$n$~ even.
Furthermore, the case ~$n=1$~ is special and is discussed
separately. (The case ~$n=2$~ is reduced to the case ~$n=1$.) Thus,
the paper is organized as follows. In Section 2 we give the general
setting. In Section 3 we specify the setting to the conformal case.
In Sections 4,5,6 we treat the cases ~$n=1,3,4$~ in detail.

\newsec{Preliminaries}

\nt Let $G$ be a  semisimple non-compact Lie group, and  $K$ a
maximal compact subgroup of $G$. Then we have an Iwasawa
decomposition ~$G=KAN$, where ~$A$~ is abelian simply connected, a
vector subgroup of ~$G$, ~$N$~ is a nilpotent simply connected
subgroup of ~$G$~ preserved by the action of ~$A$. Further, let $M$
be the centralizer of $A$ in $K$. Then the subgroup ~$P_0 ~=~ M A
N$~ is a minimal parabolic subgroup of $G$. A parabolic subgroup ~$P
~=~ M' A' N'$~ is any subgroup of $G$ (including $G$ itself) which
contains a minimal parabolic subgroup.\foot{The number of
non-conjugate parabolic subgroups is ~$2^r$, where $r=\rank\,A$,
cf., e.g., \War.}

The importance of the parabolic subgroups comes from the fact that
the representations induced from them generate all (admissible)
irreducible representations of $G$ \Lan. For the classification of
all irreducible representations it is enough to use only the
so-called {\it cuspidal} parabolic subgroups ~$P=M'A'N'$, singled
out by the condition that ~rank$\, M' =$ rank$\, M'\cap K$
\Zhea,\KnZu, so that $M'$ has discrete series representations \Har.
However, often induction from non-cuspidal parabolics is also
convenient and we shall use it below.

Let ~$\nu$~ be a (non-unitary) character of ~$A'$, ~$\nu\in\ca'^*$,
let ~$\mu$~ fix an irreducible representation ~$D^\mu$~ of ~$M'$~ on
a vector space ~$V_\mu\,$.

  We call the induced
representation ~$\chi =$ Ind$^G_{P}(\mu\otimes\nu \otimes 1)$~ an
~{\it elementary representation} of $G$ \DMPPT. (These are called
{\it generalized principal series representations} (or {\it limits
thereof}) in \Knapp.) Their spaces of functions are: \eqn\fun{
\cc_\chi ~=~ \{ \cf \in C^\infty(G,V_\mu) ~ \vr ~ \cf (gman) ~=~
e^{-\nu(H)} \cdot D^\mu(m^{-1})\, \cf (g) \} } where ~$a= \exp(H)\in
A'$, ~$H\in\ca'\,$, ~$m\in M'$, ~$n\in N'$.

For our purposes we need to restrict to ~{\it maximal}~ parabolic
subgroups ~$P$, 
(so that $\rank\,A'=1$), that may not be cuspidal
(the importance of such occurrences is explained on the example of
$\cg = so(n,2)$ below).
For the representations that we consider the character ~$\nu$~ is
parameterized by a real number ~$d$, called the conformal weight or
energy (the latter for reasons that will become clear below).

We   restrict also to the case of finite-dimensional (nonunitary)
representations ~$\mu$~    of ~$M'$.

An important ingredient in our considerations are the ~{\it unitary
lowest weight representations}~ of ~$\cg$. These can be realized as
factor-modules of Verma modules ~$V^\L$~ over ~$\cg^\bac$, where
~$\L\in (\ch^\bac)^*$, ~$\ch^\bac$ is a Cartan subalgebra of
~$\cg^\bac$, the lowest weight ~$\L = \L(\chi)$~ is determined
uniquely from $\chi$ \Dob. Unitarity means positivity w.r.t. the
Shapovalov form in which the conjugation is the one singling out
$\cg$ from $\cg^\bac$.

Actually, since our ERs are induced from finite-dimensional
representations of ~$\cm'$~ the Verma modules are always reducible.
Thus, it is more convenient to use ~{\it generalized Verma modules}
~$\tV^\L$~ such that the role of the lowest weight vector $v_0$ is
taken by the finite-dimensional space ~$V_\mu\,v_0\,$. For the
generalized Verma modules (GVMs) the reducibility is controlled only
by the value of the conformal weight $d$. Matters are arranged so
that there is a real number ~$d_0$~ called the ~{\it first reduction
point} (FRP), such that for ~$d>d_0$~ the GVMs are irreducible and
unitary. For ~$d=d_0$~ the GVM is reducible (in general) with
invariant subspace ~$I^\L$~ so that the factor space ~$L_\L \cong
\tV^\L/I^\L$~ is irreducible and unitary. For ~$d<d_0$~ in some
cases there is a discrete set of ~$d$-values for which the GVM is
reducible and again the factor space ~$L_\L$~ is irreducible and
unitary. This picture was known for the conformal cases
~$\cg=so(n,2)$~ when ~$n=3$ \FF{} (for more modern exposition cf.
also \Dobso) and ~$n=4$ \Mack{}, but was established for all
algebras with lowest/highest weight modules in \EHW.\foot{Note that
EHW \EHW{} work with highest weight modules, thus, their ranges are
limited from above, while we work with lowest weight modules and our
ranges are limited from below. The latter is done to have the
intuitive picture of positive energy spectrum bounded from below.
There is also a shift of the initial points - in our case $d$ - as
energy - is positive, while in the notation of \EHW{}  the spectrum
includes the point zero.}

We turn now to the relation of  discrete series representations with
lowest weight modules.

The unitary lowest weight generalized Verma modules are
infinitesimally equivalent to ~{\it holomorphic discrete series}
when ~$d = d_0+kc_0\,$, ~$k=A(\l_0)+1,2,\ldots$,
~$c_0,A(\l_0)\in\bbn$,  \EHW. ~The GVMs with ~$d = d_0+c_0A(\l_0)$~
are infinitesimally equivalent to the so-called ~{\it limits of
discrete series} \EHW. The latter are not related to
finite-dimensional representations.

The irreps for ~$d> d_0$~    are also called analytic continuation
of the discrete series.

In order to be more specific, in the next sections we consider the
conformal cases.

\np

\newsec{Conformal groups}

\nt Let ~$G=SO_o(n,2)$. We shall consider first the case ~$n>2$.
Then $G$ is the conformal group in ~$n$-dimensional Minkowski
space-time. The Lie algebra, i.e., the conformal algebra ~$\cg =
so(n,2)$~ has three non-trivial non-conjugate parabolic subalgebras
~$\cp_i = \cm_i \oplus \ca_i \oplus \cn_i\,$, $i=0,1,2$, where:
\eqna\parab
$$\eqalignno{ &\cm_0 \cong so(n-2)\ , \quad \ca_0\cong so(1,1) \oplus so(1,1)
\ , \quad \cn_0 \cong \bbr^{2n-2} &\parab a\cr &\cm_1 \cong
so(n-2)\oplus so(2,1)\ , \quad \ca_1\cong so(1,1) \ , \quad \cn_1
\cong \bbr^{2n-3} &\parab b\cr &\cm_2 \cong so(n-1,1)\ , \quad
\ca_2\cong so(1,1) \ , \quad \cn_2 \cong \bbr^{n} &\parab c\cr}$$
The parabolic ~$\cp_0$~ is a minimal one, and thus is cuspidal. The
other two parabolics are maximal, ~$\cp_1$~ is also cuspidal, while
~$\cp_2$~ is cuspidal only if ~$n$~ is odd.

We shall use representations induced from the parabolic ~$\cp_2$~
since the sets of finite-dimensional (nonunitary) representations of
~$\cm_2$~ are in 1-to-1 correspondence with the finite-dimensional
(unitary) representations of ~$so(n)$~ which is the semi-simple
subalgebra of the maximal compact subalgebra ~$\ck ~=~ so(n) \oplus
so(2)$. Thus, these induced representations are representations of
finite $\ck$-type \HC. 
Relatedly, the number of ERs in the corresponding multiplets is
equal to ~$\vr W(\cg^\bac,\ch^\bac)\vr\, /\, \vr
W(\ck^\bac,\ch^\bac)\vr ~=~ 2(1+\hh)$, ~$\hh\equiv [\han]$, where
~$\ch$~ is a Cartan subalgebra of both ~$\cg$~ and ~$\ck$. Note also
that ~$\ck^\bac \cong\cm_2^\bac \oplus \ca_2^\bac$. 

The Bruhat decomposition \Bru{} of $\cg$ which corresponds to this
parabolic is: \eqn\bruh{ \cg ~=~ \tcn_2  \oplus \cm_2 \oplus \ca_2
\oplus \cn_2\ ,} where ~$\tcn_2 ~=~ \th\cn_2\,$, ($\th$ is the
Cartan involution in $\cg$). The subalgebras in this decomposition
have direct physical meaning: ~$\cm_2$~ is the Lorentz algebra of
$n$-dimensional Minkowski space-time $M^n$, (the latter differs from
$\bbr^n$ by the Lorentzian metric), ~$\cn_2$~ is the translation
algebra of $M^n$, ~$\ca_2$~ is the subalgebra of dilatations,
~$\tcn_2$~ is the subalgebra of special conformal transformations of
$M^n$.

We label   the signature of the ERs of $\cg$   as follows:
\eqn\sgnd{\eqalign{ \chi ~~=&~~ \{\, n_1\,, \ldots,\, n_{\hh}\,;\,
c\, \} \ , \qquad n_j \in \bbz/2\ , \quad c=d-\han\ , \cr & \vr n_1
\vr < n_2 < \cdots <  n_{\hh}\ , \quad n ~{\rm even}\ ,\cr & 0 < n_1
< n_2 < \cdots <  n_{\hh} \ , \quad n ~{\rm odd}\ ,}}
where the last
entry of ~$\chi$~ labels the characters of $\ca_2\,$, and the first
$\hh$ entries are labels of the finite-dimensional nonunitary irreps
of $\cm_2\,$, (or of the finite-dimensional unitary irrep of
~$so(n)$), which also fulfil the requirement that the $n_i$'s are
either all integer, or all half-integer.

The reason to use the parameter ~$c$~ instead of ~$d$~ is that the
parametrization of the ERs in the multiplets is given in a simple
 intuitive way:
\eqna\sgne
$$\eqalignno{
\chi^\pm_1 ~~=&~~ \{ \eps  n_1\,, \ldots,\, n_\hh \,;\, \pm n_{\hh+1} \}
\ ,   \quad n_\hh < n_{\hh+1}\ ,
&\sgne{}\cr
\chi^\pm_2 ~~=&~~ \{ \eps  n_1\,, \ldots,\, n_{\hh-1}\,,\,  n_{\hh+1}\,;\,
\pm n_\hh \}     \cr
\chi^\pm_3 ~~=&~~ \{ \eps  n_1\,, \ldots,\, n_{\hh-2}\,,\, n_{\hh}\,,\,
n_{\hh+1}\,;\, \pm n_{\hh-1} \}     \cr
... \cr
\chi^\pm_{\hh} ~~=&~~ \{ \eps n_1\,, n_3\,,
\ldots,\, n_{\hh}\,,\, n_{\hh+1}\,;\, \pm  n_2 \}     \cr
\chi^\pm_{\hh+1} ~~=&~~ \{ \eps n_2\,, \ldots,\, n_{\hh}\,,\,
n_{\hh+1}\,;\, \pm  n_1 \}     \cr
\eps ~~=&~~ \cases{ \pm\,, ~&~ for ~$n$ ~ even \cr
                     1,  ~&~  for ~$n$ ~ odd \cr} \cr} $$
($\eps = \pm$~ is correlated with $\chi^\pm$).

The ERs in the multiplet are related by intertwining integral and
differential operators. The  integral operators were introduced by
Knapp and Stein \KnSt{}. These operators intertwine the pairs
~$\tcc^+_i$~ and ~$\tcc^-_i\,$~: \eqn\ackin{
G^+_i ~:~ \tcc^-_i \lra \tcc^+_{i} 
\,, \qquad G^-_i ~:~ \tcc^+_i \lra \tcc^-_{i}  \ , \quad i ~=~
1,\ldots,1+\hh \ . }

Matters are arranged so that in every multiplet only the ER with
signature ~$\chi^-_1$~ contains a finite-dimensional nonunitary
subrepresentation in  a finite-dimensional subspace ~$\ce$. The
latter corresponds to the finite-dimensional unitary irrep of
~$so(n+2)$~ with signature ~$\{ n_1\,, \ldots,\, n_\hh \,, \,
n_{\hh+1} \}$. The subspace ~$\ce$~ is annihilated by the operator
~$G^+_1\,$,\ and is the image of the operator ~$G^-_1\,$.

Analogously, in every multiplet only the ER with signature
~$\chi^+_1$~ contains holomorphic discrete series representation. In
fact, it contains also the conjugate anti-holomorphic discrete
series. The direct sum of the holomorphic and the antiholomorphic
representations are realized in  an invariant subspace ~$\cd$~ of
the ER ~$\chi^+_1\,$. That subspace is annihilated by the operator
~$G^-_1\,$,\ and is the image of the operator ~$G^+_1\,$.

Note that the corresponding lowest weight GVM is infinitesimally
equivalent only to the holomorphic discrete series, while the
conjugate highest weight GVM is infinitesimally equivalent to the
anti-holomorphic discrete series.

The \idos\ correspond to non-compact positive roots of the root
system of ~$so(n+2,\bbc)$, cf. \Dob. [In the current context,
compact roots of $so(n+2,\bbc)$ are those that are roots also of the
subalgebra $so(n,\bbc)$, the rest of the roots are non-compact.] Let
us denote by ~$\tcc^\pm_i$~ the representation space with signature
~$\chi^\pm_i\,$. The \idos\ act as follows: \eqn\acki{\eqalign{ &d_i
~:~ \tcc^-_i \lra \tcc^-_{i+1} \,, \quad i = 1,...,\hh \,, \quad
\forall n \cr &d'_i ~:~ \tcc^+_{i+1} \lra \tcc^+_{i} \,, \quad i =
1,...,\hh\,,  \quad \forall n \cr &d_{\hh} ~~=~~ d'_{\hh} \,, \quad
n ~ {\rm even} \cr &d_{\hh+1} ~:~ \tcc^-_{\hh+1} \lra \tcc^+_{\hh}
\,, \quad n ~ {\rm even} \cr &d_{\hh+1} ~:~ \tcc^-_{\hh} \lra
\tcc^+_{\hh+1} \,, \quad n ~ {\rm even} \cr }} The degrees of these
\idos\ are given just by the differences of the ~$c$~ entries:
\eqn\dgr{ \eqalign{ &\deg d_i ~~=~~  \deg d'_i ~~=~~ n_{\hh+2-i} -
n_{\hh+1-i} \,, \quad i ~=~ 1,\ldots,\hh \,, \quad \forall n \cr
&\deg d_{\hh+1} ~~=~~ n_2 + n_1   \,, \quad n ~ {\rm even} \cr }}
The multiplets can be seen pictorially in \Dobsrni.\foot{Actually
the diagrams in \Dobsrni{} are for the corresponding Euclidean cases
$so(n+1,1)$ -  the diagrams are the same for both signatures.} The
equalities between some \idos\ for $n$ even in \acki\ is due to the
fact that these operators are produced by singular vectors
corresponding to the same positive roots of the root system of
~$so(n+2,\bbc)$, cf. \Dob.

More explicitly, for ~$n$-even, ~$n=2\ell$, the root system of
~$so(n+2,\bbc)$~ may be given by vectors ~$e_i \pm e_j\,$, ~$\ell+1\geq i>j \geq 1$,
where ~$e_i$~ form an orthonormal basis in $\bbr^{\ell+1}$, i.e., ~$(e_i,e_j) = \d_{ij}\,$.
The non-compact roots may be taken as ~$e_{\ell+1} \pm e_i\,$. The roots
~$e_{\ell+1} - e_i\,$, $2\leq i\leq \ell$, correspond to the operators ~$d_{\ell+1-i}\,$,
the roots ~$e_{\ell+1} + e_i\,$, $2\leq i\leq \ell$, correspond to the operators ~$d'_{\ell+1-i}\,$,
the roots ~$e_{\ell+1} \pm e_1\,$ correspond to the operators ~$d_{\ell}\,$, ~$d'_{\ell}\,$, resp.

For ~$n$-odd, ~$n=2\ell+1$, the root system of
~$so(n+2,\bbc)$~ may be given by vectors ~$e_i \pm e_j\,$, ~$\ell+1\geq i>j \geq 1$,
~$e_k$, ~$1 \leq k \leq \ell+1$.
The non-compact roots may be taken as ~$e_{\ell+1} \pm e_i\,$, ~$e_{\ell+1}\,$.
The roots ~$e_{\ell+1} - e_i\,$, $1\leq i\leq \ell$, correspond to the operators ~$d_{\ell+1-i}\,$,
the roots ~$e_{\ell+1} + e_i\,$, $1\leq i\leq \ell$, correspond to the operators ~$d'_{\ell+1-i}\,$.
The root ~$e_{\ell+1}$~ has a special position, for ~$n_i\in\bbn$~
it corresponds to differential  operators of degree ~$2n_i$~ which are degenerations of the integral
operators ~$G^+_i\,$. The latter phenomenon is given explicitly for $n=3$ in \Dobso.

Another parametrization of the ER/GVM is by the so-called Dynkin
labels. These are defined as follows: \eqn\dynk{ m_i \equiv (\L+\r,
\a^\vee_i )\ ,} where ~$\L = \L(\chi)$, ~$\r$ is half the sum of the
positive roots of ~$\cg$, ~$\a_i$~ denotes the simple roots of
~$\cg$, ~$\a^\vee_i \equiv 2\a_i/(\a_i,\a_i)$~ is the co-root of
~$\a_i\,$. Often it is convenient to consider the so-called
Harish-Chandra parameters: \eqn\dynhc{ m_\a \equiv (\L+\r, \a^\vee
)\ ,} where $\a$ is any positive root of $\cg$. These parameters are
redundant, since obviously they can be expressed in terms of the
Dynkin labels, however, as we shall see below, some statements are
best formulated in their terms.

The numbers ~deg$\,d_i\,$, ~deg$\,d'_i\,$, are actually part of the
Harish-Chandra parameters which correspond to the non-compact
positive roots of ~$so(n+2,\bbc)$. From these, only ~deg$\,d_1\,$,~
corresponds to a simple root, i.e., is a Dynkin label.

These labellings will be used in the examples ~$n=3,4$~ below. 

Above we restricted to ~$n>2$. The case  ~$n=2$~ is reduced to
~$n=1$~ since ~$so(2,2) \cong so(1,2) \oplus so(1,2)$. The case
~$n=1$~ is special   and is treated separately in the next Section.

\np

\newsec{SL(2,R)}

\nt We start with the conformal case for ~$n=1$: ~$G=SO_o(1,2) \cong
SL(2,\bbr)/\bbz_2\,$. This is treated separately since ~$r=1$~ and
there is only one non-trivial parabolic. Next, the Lie algebra
~$so(1,2) \cong sl(2,\bbr)$~ is maximally split and the subalgebra
~$\cm$~ is trivial. Furthermore, this case is simpler and we can
give more details.

\newsubsec{Discrete series and limits thereof}

\nt We consider  $G=SL(2,\bbr)$ following in this Subsection Gelfand
et al   (the original results are in \GeNa,\Barg). ~The elementary
representations of $SL(2,\bbr)$ are parametrized by a complex number
$s$ and a signature $\eps = 0,1$. We shall denote the ERs by
~$D_\chi\,$, ~$\chi=[s,\eps]$.  The discrete unitary series are
realized as subspaces of ~$D_{-s} \equiv D_\chi\,$, ~with ~$\chi =
[-s,\eps = s(\mod 2)]$, when ~$s\in\bbz_+\,$. More precisely, there
are two discrete series UIRs ~$F^\pm_{-s}$, which are invariant
subspaces of ~$D_{-s}\,$.

The UIRs realized in ~$F^+_{-s}$~ nowadays are called ~{\it
holomorphic discrete series}, they have a lowest weight vector, the
UIRs realized in ~$F^-_{-s}$~ nowadays are called ~{\it
antiholomorphic discrete series}, they have a highest weight vector.
Their direct sum ~$F_{-s} =  F^+_{-s}\oplus F^-_{-s}$~ is an
invariant subspace of ~$D_{-s}$.

For $s\neq 0$ the Casimir (the infinitesimal character) of ~$D_{-s}$~ is equal to the Casimir of
~$D_{s}\,$, which contains as subrepresentation the
finite-dimensional (non-unitary) representation ~$E_s$~ of $SL(2,\bbr)$ of dimension ~$s$.
The UIRs ~$F^\pm_{0}$~ are not related to finite-dimensional representations,
correspondingly, they do not fulfil Harish-Chandra's criterion for discrete series,
and thus, nowadays they are called ~{\it limits of discrete series}.

  For fixed  ~$s\neq 0$~  the ERs  ~$D_{-s}$~
and ~$D_{s}$~  are partially equivalent, which is realized by two
  integral operators [later introduced in \KnSt{} in general]:
~$\ca_{\pm s} ~:~ D_{\mp s} \lra D_{\pm s}\,$, cf. \ackin. Thus,
these two ERs form a doublet. The operator ~$\ca_s$~ annihilates
~$F_{-s}$~ and its image is ~$E_s\subset D_s\,$. The operator
~$\ca_{-s}$~ annihilates ~$E_{s}$~ and its image is ~$F_{-s}\,$.
Note also that the factor space ~$F_s = D_s/E_s$~ is the direct sum
of two further subspaces: ~$F_{s} = F^+_{s}\oplus F^-_{s}\,$, and
that the operator ~$\ca_{-s}$~ maps ~$F^\pm_{s}$~ onto
~$F^\pm_{-s}\,$.\nl For ~$s=0$~ one has: ~$D_0 = F_{0} =
F^+_{0}\oplus F^-_{0}\,$, (setting ~$E_0=0$).

\newsubsec{Lowest weight representations}

\nt As we mentioned the  lowest
weight representations are conveniently realized via the lowest
weight Verma module ~$V^\L$~ over the complexification ~$\cg^\bac
= sl(2,\bbc)$~ of the Lie algebra ~$\cg = sl(2,\bbr)$,
and the weight ~$\L=\L(\chi)=\L(s)$~ is determined uniquely by
~$\chi\,$.

For ~$s\in\bbn$~  the Verma module ~$V^{\L(s)}$~  is reducible, it
has a finite dimensional factor-representation ~$E^+_s
=V^{\L(s)}/I^{\L(s)}$~ of real dimension $s$, (which naturally can
be identified with $E_s$ from above).

By introducing the conjugation singling out $\cg$ we can use the
Shapovalov form on ~$U(\cg^+)$, ($\cg^+$ are the raising generators
of ~$\cg^\bac$), to define a scalar product in ~$V^\L$~ and then
positivity produces the list of lowest weight modules.

Using as parameter ~$d=-s$~ the condition for positivity is ~$d\geq
-1$. The point ~$d=-1$~  is the first reduction point, which happens
in the Verma module ~$V^{\L(1)}$, cf. above. There the factor-UIR is
trivial (being one-dimensional). For ~$d >-1$~  the Verma modules
~$V^{\L(-d)}$~ are irreducible  and unitarizable. They are called
positive energy representations,  the parameter ~$d$~ being the
energy or conformal weight. They are also called analytic
continuation of the discrete series ~$F^+_{-s}\,$. In particular,
for ~$d=-s\in\bbn$~ the Verma module ~$V^{\L(-d)}$~ is infinitesimally equivalent to
the holomorphic discrete series irrep ~$F^+_{-s}\,$, while for
~$d=0$~ it is infinitesimally equivalent to the limit of holomorphic
discrete series irrep ~$F^+_{0}\,$. All this is illustrated in Fig.1. below:

A similar construction exists if we take the conjugate highest
weight Verma modules, the role of ~$F^+_{-s}\,,E^+_s$ ~ being played
by ~$F^-_{-s}\,,E^-_s$. In that case we are speaking about ~{\it
negative energy representations}, with parameter ~$d=s$, so that we
have ~$d\leq 1$.

\fig{}{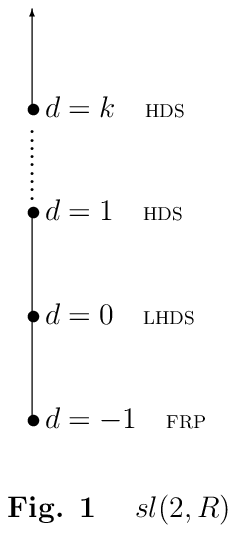}{16cm}

\np

\newsec{SO(3,2)}

\nt The algebra $so(3,2)$ has, besides the minimal parabolic, two
maximal (cuspidal) parabolic subalgebras which are isomorphic
(though non-conjugate!). So below we fix one of these maximal
parabolic subalgebras, and denote it ~$\cp_{\rm max}\,$. Here the
reducible ERs that have finite $\ck$-type representations can be
induced from ~$\cp_{\rm max} = \cm_2 \ca_2 \cn_2\,$, where ~$\cm_2 =
so(2,1)$, ~$\dim \ca_2=1$, ~$\dim \cn_2=3$. Their signatures  are
given by ~$\chi = [E_0,s_0]$, where we have introduced the
traditionally used energy $E_0$ and  spin $s_0\,$,
$s_0=0,\ha,1,\ldots$~ (the latter parametrizing the
finite-dimensional irreps of ~$so(2,1)$ or $so(3)$).

Alternatively, the ERs (GVMs) are determined by the two Dynkin
labels: ~$m_i = \lg \L+\r, \a_i^\vee \rg$, ~$i=1,2$, where $\a_i$ are the
simple roots. The relation between the two parametrisations is:
\eqn\eshc{\eqalign{& m_1 = 2s_0 +1,  \quad m_2 ~=~ 1 - E_0 - s_0\, ,
\cr & \chi = [E_0,s_0] = [\ha (3-m_1) - m_2  \,,\ha (m_1-1)]}} The
numbers ~$m_{3} = m_1+2m_2 ~=~ 3 - 2E_0 \,$,   $m_{4} = m_1+m_2 ~=~
2 - E_0 + s_0\, \,$  correspond to the two non-simple positive roots
~$\a_3= \a_1+\a_2\,$,  ~$\a_4= 2\a_1+\a_2\,$.
  The root ~$\a_1$~ is ~$\cm_2$-compact, the other roots are 
~$\cm_2$-non-compact. The set of the four numbers ~$m_i\,$, ~$i=1,2,3,4$, are
the Harish-Chandra parameters.

\newsubsec{Multiplets, finite-dimensional irreps and discrete series}

\nt
The reducible such ERs are grouped in quartets, doublets and singlets \Dobso.

The quartets are depicted in Fig. 2, cf. \Dobso{}.
The signatures of the ERs of the quartet are given by:
\eqn\signadsq{\chi^\pm_{q,k} = [\ha(3\pm(q+2k))\,, \ha (q-1)] \ ,
\quad \chi'^\pm_{q,k} = [\ha(3\pm q)\,, \ha (q-1+2k)]\ , \quad
q,k\in\bbn } The quartets are in 1-to-1 correspondence with the
finite-dimensional irreps of $G$ since in each quartet there is
exactly one ER which contains (as subrepresentation) a
finite-dimensional irrep.
The latter are parametrized by the positive integer Dynkin labels
which we denote by ~$q,k$. The corresponding finite-dimensional
irrep is denoted ~$E_{q,k}\,$, it has dimension:
~$q\,k\,(q+k)\,(q+2k)/6$, \Dobso, and is contained in the ER denoted
by ~$\chi^-_{q,k}\,$.

Consequently, the quartets also hold the discrete series
representations. For fixed ~${q,k}$~ the discrete series are
contained in ~$\chi^+_{q,k}\,$, $\chi'^+_{q,k}\,$, ~- we know that
there are two distinct non-conjugate cases of discrete series
\Knapp.

  The doublets are   denoted by ~$\chi^\pm_{q}\,$,   and the expression
for their signatures can be obtained from the signatures of
~$\chi^\pm_{q,k}$~ or from ~$\chi'^\pm_{q,k}\,$~  by setting
~$k=0$, 
 i.e., \eqn\sodub{ \chi^\pm_{q} ~=~ [\ha(3\pm q)\,, \ha (q-1)] \ , ~~q\in\bbn \ .}

Note that for all pairs of ERs with signature distinguished by ~$\pm$~
the sum of the $E_0$'s of the two ERs equals 3 - the dimension of Minkowski space-time
in this case. Furthermore, such pairs   are related
by two Knapp-Stein integral operators, as in the $SL(2,\bbr)$ case,
(though, some of those operators that act from the ~'$-$'~ER to the
~'$+$'~ER degenerate into differential operators, see below).
 Thus, each doublet can be represented pictorially by any such ~$\pm$~ pair
 from the quartet.

The singlets are $\chi^s_{n} ~=~ [\trha\,, n-\ha ]$, ~$n\in\bbn$, and the
expression for their signatures can be obtained from the
signatures of ~$\chi^\pm_{q,k}\,$ by setting ~$q=2n,~k=-\ha q=-n$~
(then $\pm$ coincide).\foot{Obviously the doublets and singlets are
not related to any finite-dimensional representations.}

In  Table 1 we give the Harish-Chandra parameters $m_\a$ for all
representations that we discuss in this Section.

According to the results of Harish-Chandra the holomorphic discrete
series happen when the numbers ~$m_\a$~ are negative integers for
the ~$\cm_2$-non-compact roots. Thus, we see from the Table that the
holomorphic discrete series are contained in the ERs ~$\chi^+_{q,k}\,$.

The limits of the holomorphic discrete series happen when some of
the ~$\cm_2$-non-compact numbers ~$m_\a$~ become zero, while the rest of the
non-compact numbers ~$m_\a$ remain negative. We see that these
limits are contained in the ERs ~$\chi^+_{q}\,$, (from the
doublets).

\newsubsec{Holomorphic discrete series and lowest weight representations}

\nt Next we discuss how the lowest weight positive energy
representations fit in the multiplets, and when they are
infinitesimally equivalent to holomorphic discrete series.

We are interested in the positive energy UIRs
 of ~$\cg$~ which are given as follows \refs{\Dirac,\Fro,\Evans} \
(with $s_0\in\ha\bbz_+$):
\foot{We have adopted the notation of \Fro{}, so that $D(E_0,s_0)$ is the
UIR which is contained as subrepresentation of the ER/GVM with signature
~$\chi = [E_0,s_0]$.}
\eqna\unita
$$\eqalignno{
& {\rm Rac} ~:~ D(E_0,s_0) ~=~ D(1/2,0) ~, \qquad {\rm Di} ~:~
D(E_0,s_0) ~=~ D(1,1/2) ~, &\unita a\cr
&D(E_0 \geq 1 , ~s_0\, =\, 0)~, 
\quad D(E_0 \geq 3/2 , ~s_0 \, =\, 1/2) ~, 
\quad D(E_0 \geq s_0 + 1 , ~s_0\,\geq\, 1) .\qquad  &\unita b}$$
The UIRs in \unita{a} are the two singleton representations discovered by Dirac
\Dirac{} and the last case in \unita{b}    corresponds to
the spin-$s_0$~ massless representations \Fro.  We note that for
these UIRs ~$m_2$~ is never a positive integer, ~$m_3$~ is a
positive integer only for ~$E_0 ~=~ 1/2, 1$, in which case ~$m_3 ~=~
2,1$, (Rac,Di), respectively. Similarly, ~$m_4$~ is a positive
integer only for ~$E_0 -s_0 ~=~ 1$, and that integer is ~$m_4 ~=~
1$.

We want to see which positive energy irreps would fit in our
multiplets. First we note that no such irreps can fit ~$\chi^-_{q,k}$~ since ~$E_0 = \trha -\ha q -k \leq 0$,
and ~$\chi'^-_{q,k}$~ since ~$E_0 = \trha -\ha q \leq 1$ but ~$s_0 = \ha (q-1) + k \geq 1$.

We notice that all positive energy irreps from \unita{b} that would fit some multiplet
can be parametrized with one parameter ~$k\in\bbz$, ~$k\geq -1$.
The parametrization is as follows:
$$ m_1 = q = 2s_0+1 \ ,\quad m_2 = -q-k \qquad \Longrightarrow \qquad E_0 = 2+s_0+k $$
In fact, for ~$k\geq 1$~ we obtain tautologically all ERs
~$\chi^+_{q,k}$, ~$q=2s_0+1$, (cf. the Table), which contain all
holomorphic discrete series. For ~$k=0$~ we obtain all ERs
~$\chi^+_{q}$, ~$q=2s_0+1$, (cf. the Table), which contain all
limits of holomorphic discrete series. Finally, for ~$k=-1$~ we
obtain the positive energy irreps ~$D(1,0)$, ~$D(\trha,\ha)$,
~$D(s_0+1,s_0)$ ($s_0\geq 1)$, which are contained in the ERs
~$\chi^-_{1}\,$, ~$\chi^s_{1}\,$, ~$\chi'^+_{2s_0-1,1}\,$.

From this we see that the above would   fit the EHW
\EHW{} picture with ~$A(\l_0) =1$, (the parameter
$z$ from \EHW{} corresponds to our $-k$). Accordingly the irreps ~$D(1,0)$,
~$D(\trha,\ha)$, ~$D(s_0+1,s_0)$ ($s_0\geq 1)$,~ are in GVMs which are FRPs.
In fact, exceptionally, the UIRs ~$D(1,0)$, ~$D(\trha,\ha)$, are isomorphic
to the corresponding GVMs since they happen to be irreducible.\foot{Brief
explanation: In the case ~$D(1,0)$~ the
corresponding Verma module is reducible under roots ~$\a_1,\a_3,\a_4\,$,
($m_1=m_3=m_4=1$), however, the singular vectors corresponding to
the non-compact roots ~$\a_3,\a_4\,$~  turn out to be descendants of
the singular vector corresponding to the compact root ~$\a_1\,$.
Consequently, when factorizing the Verma module the resulting GVM is
irreducible. In the case ~$D(\trha,\ha)$~ the  Verma module is reducible under roots ~$\a_1,\a_4\,$,
($m_1=2,m_4=1$), however, the singular vector corresponding to
the non-compact root ~$\a_4\,$~  turns out to be descendant of
the singular vector corresponding to the compact root ~$\a_1\,$.
See the details in \Dobso.}

Finally, from the list of positive energy irreps, it remains to
 discuss the Rac and the Di from \unita{a}. They are found in doublets.
 With respect to the positive energy spectrum
they are isolated points below - by $\ha$-spacing - the FRPs ~$D(1,0)$, ~$D(\trha,\ha)$, resp.

The Rac is in the reducible ER    ~$\chi^-_{1,\ha}\,$. This ER is partially
equivalent to the ER ~$\chi^+_{1,\ha}\,$ (denoted in the Table as Rac*). The latter's GVM is
irreducible and also of positive energy: ~$s_0=0$, ~$E_0=\fha$. The
intertwining operator acting from the Rac ER to Rac* is the d'Alembert operator
(obtained by reduction of a Knapp-Stein operator \Dobso).

The Di is in the reducible ER    ~$\chi'^-_{1,\ha}\,$. This ER is
partially equivalent to the ER ~$\chi'^+_{1,\ha}\,$ (denoted in the
Table as Di*). The latter's GVM is irreducible and also of positive
energy: ~$s_0=\ha$, ~$E_0=2$.

All this is illustrated on Figures 3,4.

\np

\newsec{SU(2,2)}

\nt In the 4D conformal case the   ERs/GVMs that have finite
$\ck$-type representations can be induced from the maximal parabolic
subalgebra ~$\cp_{\rm max} = \cm_2 \ca_2 \cn_2\,$, where ~$\cm_2 =
so(3,1)$, ~$\dim \ca_2=1$, ~$\dim \cn_2=4$. The signatures  are
given by ~$\chi = [j_1,j_2;d]$, where ~$j_1,j_2$~ parametrize the
finite-dimensional irreps of ~$so(3,1)$, and ~$d=2+c$~ is the energy or
conformal weight (while ~$c$~ is parametrizing the $\ca_2$-character).\foot{The ERs
of the group $G=SU(2,2)$ are induced from the parabolic subgroup
~$P_2=M_2A_2N_2\,$, cf. \fun{}, and the signature of ~$M_2 \cong
SL(2,\bbc) \times\!\!\!| \bbz_2\,$ is given by $(j_1,j_2)$
parametrizing the irreps of $sl(2,\bbc)\cong so(3,1)$ and the
signature ~$\eps = 0,1$ parametrizing the character of $\bbz_2\,$.
In what follows we shall suppress $\eps$ since in the multiplet
points (see below) it is fixed: $\eps = (p+\nu+n-1)$(mod2), and
thus, we shall consider only the Lie algebra picture.}

Alternatively, the ERs/GVMs  are determined by the three Dynkin
labels: ~$m_i = \lg \L+\r, \a_i^\vee \rg$, where $\a_i$ are the simple
roots. The relation between the two parametrisations is:
\eqn\relpar{\eqalign{& m_1 = 2j_1 +1,  \quad m_2 ~=~ 1 - d - j_1 -
j_2\, , \quad m_3 = 2j_2 +1\ ; \cr &\chi = [j_1,j_2;d=2+c] = [\ha
(m_1-1), \ha (m_3-1);\ 2-m_2 -\ha(m_1+m_3)]}} The numbers $m_{12} =
m_1+m_2\,$, $m_{23} = m_2+m_3\,$, $m_{13} = m_1+m_2+m_3\,$,
correspond to the three non-simple roots in an obvious manner. The
roots ~$\a_1,\a_3$~ are $\cm_2$-compact, the other roots are $\cm_2$-non-compact. The set of
the six numbers ~$m_\a$~ are the Harish-Chandra parameters.  We also note that the numbers ~$m_1,m_3,c$~ parametrize
the weights of the maximal compact subalgebra $\ck$ and thus, are related to the
discrete series representations, see below.

\newsubsec{Multiplets, finite-dimensional irreps and discrete series}

\nt The reducible ERs/GVMs are grouped in sextets, doublets, and
singlets \PeSo,\Dobc. The sextets are depicted in Fig. 5, cf.
\PeSo,\Dobc{}.
The signatures of the ERs/GVMs of the sextet are given by:
\eqn\signads{\eqalign{ \chi^-_{p\nu n} &= [\ha(p-1), \ha(n-1) ;\
2-\nu -\ha (p+n)] \cr \chi^+_{p\nu n} &= [\ha(n-1), \ha(p-1) ;\
2+\nu +\ha (p+n)] \cr \chi'^-_{p\nu n} &= [\ha(p+\nu-1),
\ha(n+\nu-1) ;\ 2 -\ha (p+n)] \cr \chi'^+_{p\nu n} &= [\ha(n+\nu-1),
\ha(p+\nu-1) ;\ 2 +\ha (p+n)] \cr \chi''^-_{p\nu n} &= [\ha(\nu-1),
\ha(p+n+\nu-1) ;\ 2 +\ha (p-n)] \cr \chi''^+_{p\nu n} &=
[\ha(p+n+\nu-1), \ha(\nu-1)  ;\ 2 +\ha (n-p)] }}
The sextets are in 1-to-1 correspondence with the
finite-dimensional irreps of $G$ since in each sextet there is exactly one
ER/GVM which contains (as subrepresentation) a finite-dimensional
irrep. The latter are parametrized by the positive integer Dynkin
labels which we denote as in \DoPea{} by ~$p,\nu, n\in\bbn$.
Correspondingly, the finite-dimensional irrep is denoted ~$E_{p\nu
n}$, it has dimension: ~$p\nu n (p+\nu) (n+\nu) (p+\nu+ n)/12$,
\Dobc, and is contained in the ER/GVM denoted by ~$\chi^-_{p\nu n}$.

As a consequence, the sextets also hold the discrete series
representations. For fixed ~${p,\nu, n}$~ the six representations of
the sextet are denoted by: ~$\chi^\pm_{p\nu n}\,$, ~$\chi'^\pm_{p\nu
n}\,$, ~$\chi''^\pm_{p\nu n}\,$. We know that there are three   distinct non-conjugate cases
of discrete series \Knapp. In our setting the discrete series are contained
in the cases when ~$c>0$~: ~$\chi^+_{p\nu n}\,$, $\chi'^+_{p\nu n}\,$, $\chi''^+_{p\nu
n}\,$ ($n>p$), $\chi''^-_{p\nu n}\,$ ($n<p$), (the two cases ~$\chi''^\pm_{p\nu n}\,$~
are conjugate and count as one case).

The limits of discrete series representations are in some doublets.
The doublets are of three kinds, denoted by: ~$^1\chi^\pm_{p\nu}\,$,
~$^2\chi^\pm_{p n}\,$, ~$^3\chi^\pm_{\nu n}\,$, and the expression
for their signatures can be obtained from the signatures of
~$\chi'^\pm_{p\nu n}\,$, by setting, ~$n=0$, ~$\n=0$, ~$p=0$,
respectively, i.e.,
\eqn\signadsd{\eqalign{
^1\chi'^-_{p\nu } &= [\ha(p+\nu-1), \ha(\nu-1) ;\ 2 -\ha p] \cr
^1\chi'^+_{p\nu } &= [\ha(\nu-1), \ha(p+\nu-1) ;\ 2 +\ha p] \cr
^2\chi'^-_{p n} &= [\ha(p-1), \ha(n-1) ;\ 2 -\ha (p+n)] \cr
^2\chi'^+_{p n} &= [\ha(n-1), \ha(p-1) ;\ 2 +\ha (p+n)] \cr
^3\chi'^-_{\nu n} &= [\ha(\nu-1), \ha(n+\nu-1) ;\ 2 -\ha n] \cr
^3\chi'^+_{\nu n} &= [\ha(n+\nu-1), \ha(\nu-1) ;\ 2 +\ha n] \cr
  }}

Note that for all pairs of ERs/GVMs with signature distinguished by
~$\pm$~ the sum of the conformal weights  $d$ of the two ERs/GVMs equals 4 - the
dimension of Minkowski space-time in this case. Furthermore, the ERs
of such pairs are related by two Knapp-Stein integral operators
\KnSt.

Finally the singlets are denoted by ~$\chi^s_{\nu} ~=~
[\ha(\nu-1), \ha(\nu-1) ;\ 2 ]$, ($\nu\in\bbn$), and the
expression for their signatures can be obtained from the
signatures of ~$\chi'^\pm_{p\nu n}\,$, by setting, ~$n=0$~ and
~$p=0$ (then $\pm$ coincide).\foot{Obviously the doublets and
singlets are not related to any finite-dimensional
representations.}

In  Table 2 we give the Harish-Chandra parameters $m_\a$ for all
representations that we discuss in this Section.\foot{Matters are
arranged as discussed so that the Dynkin labels are equal to
$p,\nu,n$ for the series ~$\chi^-_{p\nu n}\,$ (which contains the
finite-dimensional irreps).}

According to the results of Harish-Chandra the holomorphic discrete
series happen when the numbers ~$m_\a$~ are negative integers for
the $\cm_2$-non-compact roots. Thus, we see from the Table that the
holomorphic discrete series are contained in the ERs ~$\chi^+_{p\nu
n}\,$. The limits of the holomorphic discrete series happens when
some of the non-compact numbers ~$m_\a$~ become zero, while the rest
of the non-compact numbers ~$m_\a$ remain negative. We see that
these limits are contained in the ERs ~$^2\chi^+_{p n}\,$ (obtained
also from ~$\chi^+_{p\nu n}\,$ for $\nu$ =0).

\newsubsec{Holomorphic discrete series and lowest weight representations}

\nt Next we discuss how the lowest weight positive energy
representations fit in the multiplets, and when they are
infinitesimally equivalent to holomorphic discrete series.

There are 2 basic cases of positive energy representations \Mack:\nl
$$ 1)  ~j_1j_2\neq 0, \qquad 2)  ~j_1j_2 = 0$$

In case 1) the positive energy representations fulfil the condition
\Mack{}:\nl
$$ d \geq 2 + j_1+j_2 \ , \quad   j_1j_2\neq 0 $$

 For ~$d > 2 +j_1+j_2$~ the GVMs are irreducible and unitary.

 The point ~$d=d_0\equiv 2 + j_1+j_2$~  is the first reduction point. In our picture it is
realized in the GVM with signature ~$\chi'^+_{p 1 n}$, so that
~$j_1=\ha n$, ~$j_2=\ha p$.

The point ~$d=d_0+1$~ is a limit of discrete series, while the
integer points with ~$d\geq d_0+2$~ are the holomorphic discrete
series. Indeed, the former  are contained in ~$^2\chi^+_{p n}\,$,
while the latter are contained in the ERs with signature ~$\chi^+_{p
\nu n}$. In both cases we have ~$j_1=\ha (n-1)$, ~$j_2=\ha (p-1)$,
~$d=2+\nu+ \ha (n+p)$, where ~$n,p>1$~ and ~$\nu=0$, $\nu\in\bbn$,
distinguishes the two cases.

Thus, these cases correspond to ~$c_0=1$~ (see above) and
~$A(\l_0)=1$~ in the terminology of \EHW{}. Here and below the
unitarity parameter ~$z$~ of \EHW{} is related to ours as: ~$z ~=~
-d + d_0 + A(\l_0)\,$.

In case 2) the positive energy representations fulfil the condition
\Mack{}:\nl
$$ d \geq 1 + j_1+j_2 \ , \quad   j_1j_2 = 0 $$

For ~$d > 1 + j_1+j_2$~ the GVMs are irreducible and unitary.

The point ~$d^0_0= 1 + j_1+j_2$~ is the first reduction point. These
are the ~{\it massless} representations of $so(4,2)$.

For ~$j_1+j_2 \geq 1$~ the FRP  is realized in the ERs/GVMs with
signatures: ~$\chi''^+_{1 1 n}$, with ~$j_1=\ha (n+1)\geq 1$, ~$j_2=
0$, ~and ~$\chi''^-_{p 1 1}$, with ~$j_1= 0$, ~$j_2=\ha (p+1)\geq
1$.

For ~$j_1+j_2 =\ha$~ the FRP is realized in the  ERs/GVMs with
signatures: ~$^3\chi^-_{1 1}$, with ~$j_1=\ha$, ~$j_2= 0$, ~and
~$^1\chi^-_{1 1 }$, with ~$j_1= 0$, ~$j_2=\ha\,$. \foot{The two
conjugated representations are two-component massless Weyl spinors.
They are partially equivalent to the ERs ~$^3\chi^+_{1 1}$,
~$^1\chi^+_{1 1}$ mentioned below, and the corresponding Knapp-Stein
operators from these FRPs degenerate to the two well-known first
order conjugated Weyl equations.}

For ~$j_1=j_2 = 0 $~ the FRP is realized in the ER with signature:
~$^2\chi^-_{11}\,$.\foot{This massless scalar
representation is partially equivalent to the scalar ER
~$^2\chi^+_{1 1}$, mentioned below. The two ERs  are related by
Knapp-Stein integral operators, however, the operator   from
~$^2\chi^-_{11}$~ to ~$^2\chi^+_{11}$~ degenerates to the d'Alembert
operator. That d'Alembert operator arises also as a conditionally
invariant differential operator due to the presence of a subsingular
vector in the corresponding Verma module with signature $(m_1,m_2,m_3) =
(1,0,1)$ \Dobcond.}

As we shall see, these cases correspond to ~$c_0=1$~ (see above).
and ~$A(\l_0)=2$~ in the terminology of \EHW{} (the FRP is
$z=A(\l_0)=2$).

The point next  to the FRP ($z=1$ by \EHW)
 with ~$d = d^0_0 +1 = 2 + j_1+j_2$~ is part of the analytic continuation of the discrete series.\nl
For ~$j_1+j_2 \geq \ha$ it fits the ERs ~$^3\chi^+_{1 n}$, so that
~$j_1=\ha n$, ~$j_2= 0$, and ~$^1\chi^+_{p 1 }$, so that ~$j_1= 0$,
~$j_2=\ha p$.\nl For ~$j_1=j_2 = 0 $~ it is realized in the singlet
ER with signature: ~$\chi^s_{1}$.

The next point with   ~$d = d^0_0 +2 = 3 + j_1+j_2$, ($z=0$ by
\EHW),    fits the ERs ~$^2\chi^+_{p n}$ with either ~$p=1$ or
$n=1$, which contain limits of discrete series (with ~$j_1=\ha (n-1)$, ~$j_2=\ha
(p-1)$, as above for $j_1 j_2  \neq 0$).

Finally, the cases with integer $d\geq d^0_0 +3 =4+ j_1+j_2$ ~($z<0$
by \EHW) are realized by the ERs ~$\chi^+_{p \nu n}$ which contain
the holomorphic discrete series (as above for $j_1 j_2 \neq 0$).

All this is illustrated on Figures 6,7.

\newsubsec{Induction from another parabolic}

\nt The ERs discussed until now can be also induced from the minimal
parabolic subgroup as shown in \DoMo, however, then they appear in
larger multiplets \Dobc. In particular, the sextets are part of
24-plets, where each ER corresponds to an element of the Weyl group
$W$ of ~$\cg^\bac = so(6,\bbc)$, (recall that ~$|W|=24$).

But besides the minimal and maximal non-cuspidal parabolics (of
dimensions 9 and 11, resp.), the group $SU(2,2)$  has a maximal
cuspidal parabolic of dimension 10: ~$P_1 = M_1 A_1 N_1\,$, where
~$M_1 = SL(2,\bbr)\times SO(2)$, ~$\dim A_1=1$, ~$\dim N_1=5$. The
signatures here are ~$\chi_1 ~=~ [n',k,\eps,\n']$, where
~$n'\in\bbz$~ is a character of ~$SO(2)$, ~$\n'\in\bbc$~ is a
character of ~$A_1\,$, ~$k,\eps$~ fix a discrete series
representation of ~$SL(2,\bbr)$, ~$k\in\bbn$, ~$\eps =0,1$, or a
limit thereof when $k=0$. In the integer points we are interested in
~$\eps = k(\mod 2)$, ~$\n' \in\bbz $, and the relation with the
Dynkin labels is as follows \Dobc: \eqn\intp{ m_1 ~=~ \ha (k-\n'
+n') \ ,\quad m_2 ~=~ -k \ , \quad m_3 ~=~ \ha (k-\n' -n')}

Clearly, representations induced in this way will not describe all
ERs with finite K-type, cf. the Table. For instance, one can not
obtain the ERs of type ~$\chi^-_{p\n n}$. Further, there are some
restrictions on the values of ~$n',k,\n'$~ when matching the other
five series, e.g., to have finite K-types must hold the condition:
~$k > \nu' + |n'|$. Furthermore, in order to fit the holomorphic
discrete series, i.e., ~$\chi^+_{p\n n}$, must hold: ~$k > |\nu'|
+ |n'|$.  These representations  describe  the limits of discrete
series when $\nu'=0$ and ~$k > |n'|$.

\np

\newsec{Outlook}

\nt In the present paper we restricted to the conformal group case.
Similar explicit descriptions can be easily achieved for the other
non-compact groups with lowest/highest weight representations. We
plan also to extend these considerations \Dobp{} to the
supersymmetric cases using precious results on the classification of
positive energy irreps in various dimensions
\DoPe,\Min,\Dobsu,\CCTV, and also to the quantum group setting using
\Dobqg. Such considerations are expected to be very useful    for
applications to string theory and integrable models, cf., e.g.,
\Witten.

\bigskip

\nt {\bf Acknowledgements.}

\nt The author would like to thank for hospitality the International
School for Advanced Studies, Trieste, where part of the work was
done. This work was supported in part by the Alexander von Humboldt
Foundation in the framework of the Clausthal-Leipzig-Sofia
Cooperation, and the European RTN 'Forces-Universe', contract
MRTN-CT-2004-005104.

The author would like to thank the referees, since their remarks
contributed to improving the exposition.

\np \fig{}{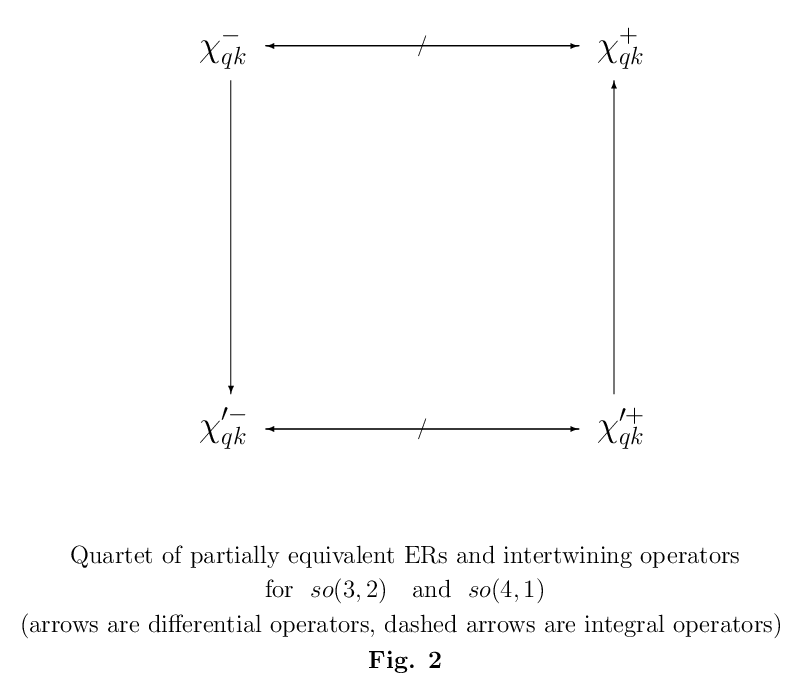}{12cm}

\np

\centerline{\bf Table 1}\medskip
\vbox{
\offinterlineskip\hrule
\vskip0.4truecm
\noindent Harish-Chandra parameters for {\bf so(3,2)}~:\nl
$m_i =  \lg \L+\r, \a_i^\vee\rg$, $i=1,2,3,4$, ~$m_{3} = m_1+2m_2\,$,   $m_{4} =  m_1+m_2 \,$, ~$m_1\equiv 2s_0+1$,
\vskip0.4truecm
\halign{\baselineskip12pt
\strut
#&
\vrule#\hskip0.1truecm &
#\hfil&
\vrule#\hskip0.1truecm &
#\hfil&
\vrule#\hskip0.3truecm &
#\hfil&
\vrule#\hskip0.1truecm &
#\hfil&
\vrule#\hskip0.1truecm &
#\hfil&
\vrule#\hskip0.1truecm &
#\hfil&
#\cr
\tablerule
&&&&  &&&& &&&& &\cr
&&\hskip 2mm ER
&&\hskip 2mm $m_1$
&&\hskip 2mm $m_2$
&&\hskip 2mm $m_3$
&&\hskip 2mm $m_4$
&&&
\cr
&&&  &&& &&& &&&&\cr
\tablerule
&&&& &&&&  &&&&&\cr
&&\hskip 2mm $\chi^-_{q,k}$
&&\hskip 2mm $q$
&&\hskip 2mm $k$
&&\hskip 2mm $q+2k$
&&\hskip 2mm $q+k$
&&&\cr
&&&&&&&&&&&&&\cr
\tablerule
&&\hskip 2mm $\chi^+_{q,k}$
&&\hskip 2mm $q$
&&\hskip 2mm $-q-k$
&&\hskip 2mm $-q-2k$
&&\hskip 2mm $-k$
&&&\cr
&&&&&&&&&&&&&\cr
\tablerule
&&\hskip 2mm $\chi'^-_{q,k}$
&&\hskip 2mm $q+2k$
&&\hskip 2mm $-k$
&&\hskip 2mm $q$
&&\hskip 2mm $q+k$
&&&\cr
&&&&&&&&&&&&&\cr
\tablerule
&&\hskip 2mm $\chi'^+_{q,k}$
&&\hskip 2mm $q+2k$
&&\hskip 2mm $-q-k$
&&\hskip 2mm $-q$
&&\hskip 2mm $k$
&&&\cr
&&&&&&&&&&&&&\cr
\tablerule
&&\hskip 2mm $\chi^-_{q}$
&&\hskip 2mm $q$
&&\hskip 2mm $0$
&&\hskip 2mm $q$
&&\hskip 2mm $q$
&&&\cr
&&&&&&&&&&&&&\cr
\tablerule
&&\hskip 2mm $\chi^+_{q}$
&&\hskip 2mm $q$
&&\hskip 2mm $-q$
&&\hskip 2mm $-q$
&&\hskip 2mm $0$
&&&\cr
&&&&&&&&&&&&&\cr
\tablerule
&&\hskip 2mm $\chi^s_{n}$
&&\hskip 2mm $2n$
&&\hskip 2mm $-n$
&&\hskip 2mm $0$
&&\hskip 2mm $n$
&&&\cr
&&&&&&&&&&&&&\cr
\tablerule
&&\hskip 2mm Rac, $\chi^-_{1,\ha}$
&&\hskip 2mm 1
&&\hskip 2mm $\ha$
&&\hskip 2mm 2
&&\hskip 2mm $\trha$
&&&\cr
&&&&&&&&&&&&&\cr
\tablerule
&&\hskip 2mm Rac$^*$, $\chi^+_{1,\ha}$
&&\hskip 2mm 1
&&\hskip 2mm $-\trha$
&&\hskip 2mm -2
&&\hskip 2mm $-\ha$
&&&\cr
&&&&&&&&&&&&&\cr
\tablerule
&&\hskip 2mm Di, $\chi'^-_{1,\ha}$
&&\hskip 2mm 2
&&\hskip 2mm $-\ha$
&&\hskip 2mm 1
&&\hskip 2mm $\trha$
&&&\cr
&&&&&&&&&&&&&\cr
\tablerule
&&\hskip 2mm Di$^*$, $\chi'^+_{1,\ha}$
&&\hskip 2mm 2
&&\hskip 2mm $-\trha$
&&\hskip 2mm -1
&&\hskip 2mm $\ha$
&&&\cr
&&&&&&&&&&&&&\cr
\tablerule
}}

\np

\fig{}{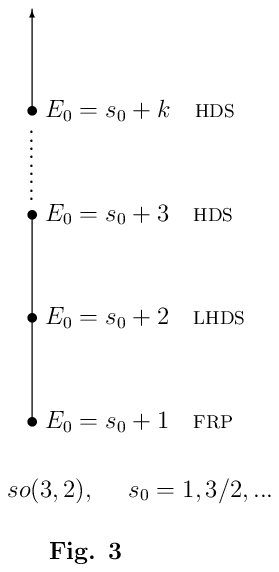}{18cm}

\np

\fig{}{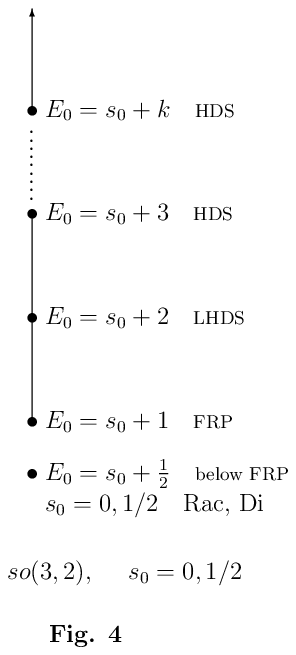}{18cm}

\np

\voffset 2cm
 \fig{}{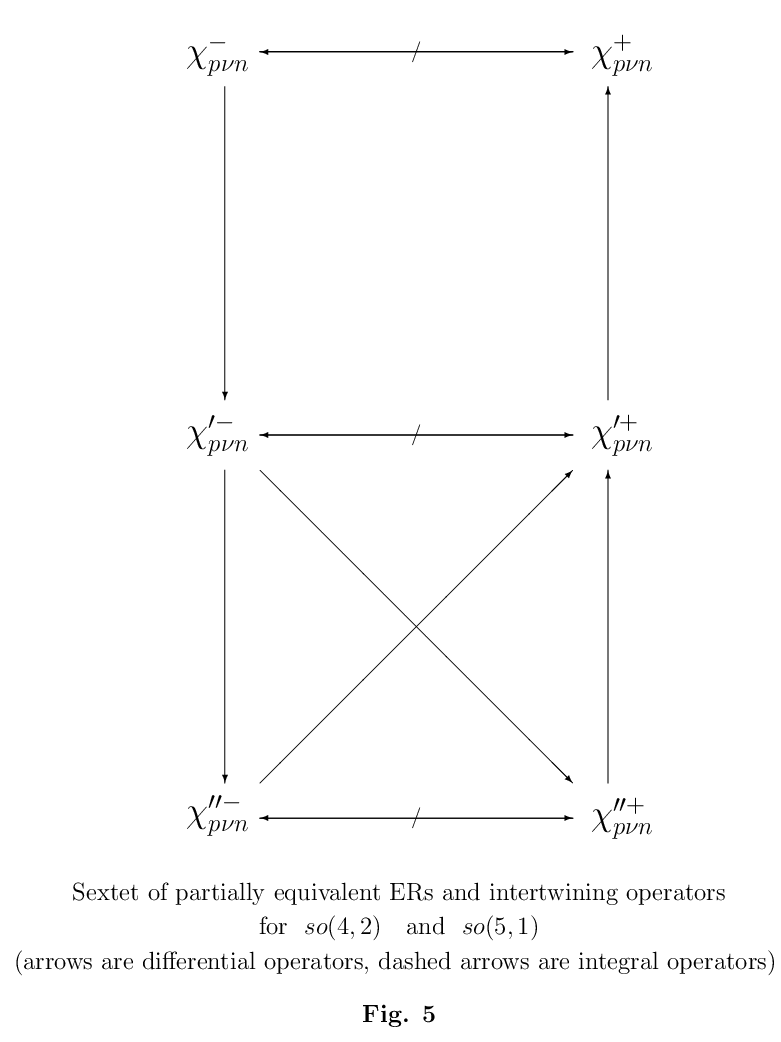}{12cm}

\np

\hoffset -1cm

\centerline{\bf Table 2}\medskip \vbox{ \offinterlineskip\hrule
\vskip0.4truecm \noindent Harish-Chandra parameters for {\bf so(4,2)} ~:\nl
$m_i = \lg\L+\r, \a_i \rg$, $i=1,2,3$, ~$m_{12} =
m_1+m_2\,$, ~$m_{23} = m_2+m_3\,$, ~$m_{13} = m_1+m_2+m_3\,$
\vskip0.4truecm \halign{\baselineskip12pt \strut #&
\vrule#\hskip0.1truecm & #\hfil& \vrule#\hskip0.1truecm & #\hfil&
\vrule#\hskip0.3truecm & #\hfil& \vrule#\hskip0.1truecm & #\hfil&
\vrule#\hskip0.1truecm & #\hfil& \vrule#\hskip0.1truecm & #\hfil&
\vrule#\hskip0.1truecm & #\hfil& \vrule#\hskip0.1truecm & #\hfil&
#\cr \tablerule &&&& &&&& &&&& &&&& &\cr &&\hskip 2mm ER &&\hskip
2mm $m_1$ &&\hskip 2mm $m_2$ &&\hskip 2mm $m_3$ &&\hskip 2mm
$m_{12}$ &&\hskip 2mm $m_{23}$ &&\hskip 2mm $m_{13}$ &&& \cr &&& &&&
&&& &&& &&&&&\cr \tablerule &&&& &&&& &&&& &&&&&\cr &&\hskip 2mm
$\chi^-_{p\nu n}$ &&\hskip 2mm $p$ &&\hskip 2mm $\nu$ &&\hskip 2mm
$n$ &&\hskip 2mm $p+\nu$ &&\hskip 2mm $n+\nu$ &&  $p+n+\nu$ &&&\cr
&&&&&&&&&&&&&&&&&\cr \tablerule &&\hskip 2mm $\chi^+_{p\nu n}$
&&\hskip 2mm $n$ &&  $-p-\nu-n$ &&\hskip 2mm $p$ &&\hskip 2mm
$-p-\nu$ &&\hskip 2mm $-n-\nu$ &&\hskip 2mm $-\nu$ &&&\cr
&&&&&&&&&&&&&&&&&\cr \tablerule &&\hskip 2mm $\chi'^-_{p\nu n}$
&&\hskip 2mm $p+\nu$ &&\hskip 2mm $-\nu$ &&\hskip 2mm $n+\nu$
&&\hskip 2mm $p$ &&\hskip 2mm $n$ &&  $p+n+\nu$ &&&\cr
&&&&&&&&&&&&&&&&&\cr \tablerule &&\hskip 2mm $\chi'^+_{p\nu n}$
&&\hskip 2mm $n+\nu$ && $-p-\nu-n$ &&\hskip 2mm $p+\nu$ &&\hskip 2mm
$-p$ &&\hskip 2mm $-n$ &&\hskip 2mm $\nu$ &&&\cr
&&&&&&&&&&&&&&&&&\cr \tablerule &&\hskip 2mm $\chi''^-_{p\nu n}$
&&\hskip 2mm $\n$ &&\hskip 2mm $-\nu-p$ &&\hskip 2mm  $p+n+\nu$
&&\hskip 2mm $-p$ &&\hskip 2mm $n$ &&\hskip 2mm $n+\nu$ &&&\cr
&&&&&&&&&&&&&&&&&\cr \tablerule &&\hskip 2mm $\chi''^+_{p\nu n}$
&&\hskip 2mm $p+n+\nu$ &&\hskip 2mm $-n-\nu$ &&\hskip 2mm $\n$
&&\hskip 2mm $p$ &&\hskip 2mm $-n$ &&\hskip 2mm $p+\nu$ &&&\cr
&&&&&&&&&&&&&&&&&\cr \tablerule

&&\hskip 2mm ${^1}\chi^-_{p\nu}$ &&\hskip 2mm $p+\nu$ &&\hskip 2mm
$-\nu$ &&\hskip 2mm $\nu$ &&\hskip 2mm $p$ &&\hskip 2mm $0$ &&\hskip
2mm $p+\nu$ &&&\cr &&&&&&&&&&&&&&&&&\cr \tablerule

&&\hskip 2mm ${^1}\chi^+_{p\nu}$
&&\hskip 2mm $\nu$
&&\hskip 2mm $-p-\nu$
&&\hskip 2mm $p+\nu$
&&\hskip 2mm $-p$
&&\hskip 2mm $0$
&&\hskip 2mm $\nu$
&&&\cr
&&&&&&&&&&&&&&&&&\cr
\tablerule

&&\hskip 2mm $^2\chi^-_{p n}$
&&\hskip 2mm $p$
&&\hskip 2mm $0$
&&\hskip 2mm $n$
&&\hskip 2mm $p$
&&\hskip 2mm $n$
&&\hskip 2mm $p+n$
&&&\cr
&&&&&&&&&&&&&&&&&\cr
\tablerule
&&\hskip 2mm $^2\chi^+_{p n}$
&&\hskip 2mm $n$
&&\hskip 2mm $-p-n$
&&\hskip 2mm $p$
&&\hskip 2mm $-p$
&&\hskip 2mm $-n$
&&\hskip 2mm $0$
&&&\cr
&&&&&&&&&&&&&&&&&\cr
\tablerule
&&\hskip 2mm $^3\chi^-_{\nu n}$
&&\hskip 2mm $\nu$
&&\hskip 2mm $-\nu$
&&\hskip 2mm $n+\nu$
&&\hskip 2mm $0$
&&\hskip 2mm $n$
&&\hskip 2mm $n+\nu$
&&&\cr
&&&&&&&&&&&&&&&&&\cr
\tablerule
&&\hskip 2mm $^3\chi^+_{\nu n}$
&&\hskip 2mm $n+\nu$
&&\hskip 2mm $-\nu-n$
&&\hskip 2mm $\nu$
&&\hskip 2mm $0$
&&\hskip 2mm $-n$
&&\hskip 2mm $\nu$
&&&\cr
&&&&&&&&&&&&&&&&&\cr
\tablerule
&&\hskip 2mm $\chi^s_{\nu}$
&&\hskip 2mm $\nu$
&&\hskip 2mm $-\nu$
&&\hskip 2mm $\nu$
&&\hskip 2mm $0$
&&\hskip 2mm $0$
&&\hskip 2mm $\nu$
&&&\cr
&&&&&&&&&&&&&&&&&\cr
\tablerule
&&  $P^1_{n',k,\eps,\n'}$ &&
$\ha(k-\n'+n')$ &&  $-k$ &&  $\ha(k-\n'-n')$ && $\ha(n'-k-\n')$ &&
$-\ha(k+\n'+n')$ &&  $-\n'$ &&&\cr &&&&&&&&&&&&&&&&&\cr \tablerule
} }

\np

\voffset -5cm
\hoffset 3cm
\fig{}{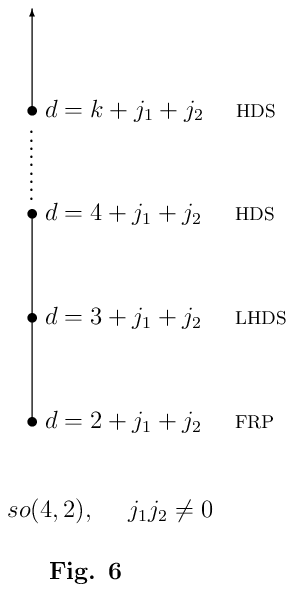}{20cm}

\np

\fig{}{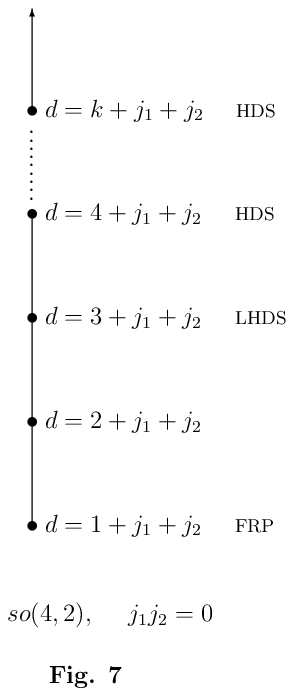}{20cm}

\np

\voffset 0cm
\hoffset 0cm
\parskip=0pt
\listrefs

\end